\documentclass{rsauthor}
\usepackage{times}
\usepackage{amsmath}
\usepackage{amssymb}
\usepackage{graphicx}
\usepackage{natbib}
\usepackage{psfrag}
 \usepackage{setspace}
 \usepackage{color}

\title{Trends, noise and reentrant long-term persistence in Arctic sea ice}

\author{ S. Agarwal$^{1,2}$, W. Moon$^{2}$, \& J. S. Wettlaufer$^{2,3,4,5}$  }

\address{$^{1}$Department of Mathematics, Indian Institute of Technology Guwahati, Guwahati 781 039, Assam, India\\$^{2}$Department of Geology and Geophysics,$^{3}$Department of Physics, and \\$^{4}$Program in Applied Mathematics, Yale University, New Haven, CT 06520-8109, USA\\$^{5}$NORDITA,  Roslagstullsbacken 23, SE-10691 Stockholm, Sweden}

\date{\today}

\abstract{
We examine the long-term correlations and multifractal properties of daily satellite retrievals of Arctic sea ice albedo and extent, for periods of $\sim$ 23 years and 32 years respectively.  The approach harnesses a recent development called Multifractal Temporally Weighted Detrended Fluctuation Analysis (MF-TWDFA), which exploits the intuition that points closer in time are more likely to be related than distant points. In both data sets we extract multiple crossover times, as characterized by generalized Hurst exponents, ranging from synoptic to decadal.  The method goes beyond treatments that assume a single decay scale process, such as a first-order autoregression, which cannot be justifiably fit to these observations.
Importantly, the strength of the seasonal cycle ``masks'' long term correlations on time scales beyond seasonal.  When removing the seasonal cycle from the original record, the ice extent data exhibits white noise behavior from seasonal to bi-seasonal time scales, whereas the clear fingerprints of the short (weather) and long ($\sim$ 7 and 9 year) time scales remain, the latter associated with the recent decay in the ice cover. Therefore, long term persistence is reentrant beyond the seasonal scale and it is not possible to distinguish whether a given ice extent minimum/maximum will be followed by a minimum/maximum that is larger or smaller in magnitude.
 
}


\jname{Proc. R. Soc. A}
\begin{document}
\maketitle

\section{Introduction}

\begin{figure}
\centering
\includegraphics[width=8.5cm]{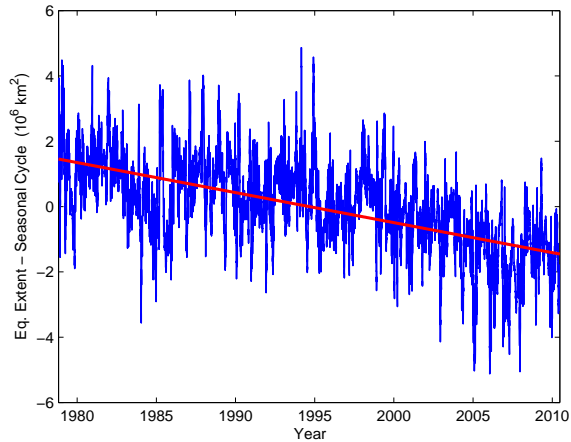}
\caption{Equivalent Ice Extent (EIE) during the satellite era, with the seasonal cycle removed, and the trend from a linear
regression shown as the red line.   EIE, which differs from traditional ice area, is  defined as the total surface area, including land, north of the zonal-mean ice edge latitude, and thus is proportional to the sine of the ice edge latitude.  EIE was defined by \citet[][]{IanGeom} to deal with the geometric muting of ice area associated with the seasonal bias of the influence of the Arctic basin land mass boundaries. Here we use EIE and hence refer to it either via this acronym or simply ice extent.}
\label{fig:coverage}
\end{figure}

Earth's polar oceans are viewed as a bellwether of climate change because their surfaces are covered by a thin (several meters) mosaic of high albedo sea ice floes that modulate the atmosphere/ocean heat flux.  Whence, as opposed to the massive meteoric ice sheets that are several kilometers thick, sea ice is considered to be a more sensitive  component of the cryosphere to perturbations and feedbacks, particularly the ice-albedo feedback which has driven large scale climate events over Earth history 
\cite[][]{Saltzman:2002}.  Indeed, the retreat of Arctic sea ice coverage during recent decades (Fig. \ref{fig:coverage})  has captured substantial interest \cite[][]{OneWatt, Serreze:2011}.   An essential question concerns the nature of the decay in ice coverage; is it a trend associated with the influence of greenhouse forcing, or is it a fluctuation in a quantitative record that is short ($\sim$ 30 years) relative to the dynamics of the cryosphere on climatic epochs ($\gtrsim$ 10$^6$ yrs)?  Our goal is to extract the longest intrinsic time scale from the data to provide a framework for geophysical modeling. 

Both past climate data and basic physical arguments indicate that a sufficiently large increase in greenhouse gas concentration will drive the decay of the ice cover, whereas the state of the art global climate models under-project the observed recent decay \cite[][]{OneWatt}.  The stabilizing response to deleterious perturbations of the ice cover was examined by \citet[][]{Tietsche:2011}, who numerically prescribed ice-free summer states at various times during the projection of 21$^{\text{st}}$ century climates and found that ice extent typically recovered within several years.  Such rapid response times can be captured within the framework of relatively simple theory \cite[][]{MW:2011}, but independent of model complexity, both internal and external forcings and their intrinsic time scales manifest themselves in large scale observations of the geophysical state of the system.  Due to the fact that we  cannot {\em a-priori} exclude the observed decline in the ice cover as being an intrinsic decadal oscillation or nonstationary influence in the climate system, we use the finest temporal resolution in the observed record (days) to examine the action of multiple scales, from that of the weather and beyond. 
The fingerprints of the noisy dynamics of the system on time scales longer than the seasonal record may reside in that record itself, and our goal here is to extract these to provide a framework for geophysical modeling of the underlying processes that created them.  It is hoped that the approach will provide theoretical constraints as well as guidelines for comparative data analysis studies.  

Most observational studies of the satellite records of ice coverage extrapolate in time the annual or monthly means \cite[e.g., see Fig. 1 of][]{Serreze:2011}.  Whilst the observed declines over this troika of decades (particularly the last decade) are striking, our goal here is to begin a systematic effort to examine the many processes that influence the ice cover viz., the intrinsic time scales reflected in geophysical data.  The general methodological framework we employ allows one to examine time series data in a manner that can distinguish between long term correlations and trends.   We examine whether there exists a multiplicity of persistent scales in the data that can provide a basis for examining cause and effect in the geophysical scale observables of the system.   Therefore, we view several types of large-scale sea ice data as a multifractal system in which a spectrum of scaling--the singularity spectrum--characterizes the behavior of nearby points  \cite[][]{Stanley:1988, Jens}.  The basic approach of relevance is the multi-fractal generalization of Detrended Fluctuation Analysis (DFA) advanced and widely applied by Kantelhardt and colleagues \cite[][]{Kantelhardt:2002, Kantelhardt:2006} aptly called Multifractal Detrended Fluctuation Analysis (MF-DFA).  In the last decade this approach has been developed in many directions, from studying extreme events with nonlinear long term memory \cite[][]{Bogachev:2011}, to examining the influence of additive noise on long term correlations  \cite[e.g.,][]{Bunde:2011}.  Here we use a new extension of this methodology  called Multifractal Temporally Weighted Detrended Fluctuation Analysis (MF-TWDFA), which exploits the intuition that in any time series points closer in time are more likely to be related than distant points  and can provide a rather more clear signature of long time scales in the fluctuation function and its moments \cite[][]{Zhou:2010}.   In the next section we summarize MF-TWDFA and describe its connection to the conventional approaches of analyzing autocorrelation functions and power spectra.  The geophysical data are described in Section \ref{sec:data} after which we present the principal results from MF-TWDFA in Section \ref{sec:results} before concluding. 

\section{Multifractal Temporally Weighted Detrended Fluctuation Analysis (MF-TWDFA) \label{sec:twdfa}}

Here, we harness a new variant of MF-DFA to study the long-range correlations and fractal scaling properties of time series data of basin wide Arctic sea ice extent and satellite albedo retrievals. The new variant, called multifractal temporally weighted detrended fluctuation analysis (MF-TWDFA), was recently developed by \citet[][]{Zhou:2010} and uses weighted regression to express the intuition that points closer in time are more likely to be similar than those that are more distant.  To insure that this paper is reasonably self contained we motivate the approach by first discussing the general rationale for multifractals versus the treatment of time series using autocorrelations and spectral analysis.  We then describe MF-DFA followed by a development of the aspects that distinguish MF-TWDFA from MF-DFA.

\subsection{Rationale for a multifractal approach \label{sec:mf}}

Consider a time series of length $i = 1, ... , N$ describing a quantity $X_i$. The linear two-point autocorrelation function $C(s)$ for realizations of $X_i$ separated by an increment $s$ is given by 
\begin{equation}
C(s) = \frac{1}{{{\sigma}_X^2} (N - s)} \sum_{i=1}^{N - s} \left(X_i - \bar{X_i} \right) \left(X_{i+s} - \bar{X_i} \right), 
\end{equation}
in which ${{\sigma}_X^2}$ is the variance and $\bar{X_i}$ is the mean.  
{\em Long-term persistence} in a time series occurs when on average the linear correlations become arbitrarily long \cite[e.g., see ][]{Jens}.  Whence, $C(s) \propto s^{-\gamma}$ and the mean correlation time ${\cal{T}}_s \equiv \int_0^N C(s) ds$ diverges as $N\rightarrow\infty$ for $0 < \gamma < 1$. Whereas, when $\gamma \ge 1$ we have finite ${\cal{T}}_s$ and hence we say that the $X_i$ are {\em short-term correlated}.   Finally, when $C(s) = 0$ for $s > 0$ then the $X_i$ are {\em uncorrelated}.  Therefore, there may be a time $s^\star$ that delimits correlated from uncorrelated $X_i$ viz.,   $C(s < s^\star) > 0$ and $C(s > s^\star) = 0$.

Despite the simplicity of characterizing a system using a single scaling exponent $\gamma$ it has long been understood that a wide range of natural and laboratory systems have correlations that cannot be captured by such an approach \cite[][]{Stanley:1988, Jens}.    It is particularly important to extract the scaling behavior for the long-term correlations, wherein the variability on long time scales may be poorly represented by a single scaling exponent and long-term trends or periodicity may obscure the calculation of the autocorrelation function.  Moreover, the distinction between trends and long-term correlations in stationary time series can become blurred.  These comprise some of the reasons for the introduction of a multifractal description of the correlations which replaces a single value with a continuous spectrum of scaling exponents.

\subsection{Multifractal detrended fluctuation analysis \label{sec:dfa}}

There are four stages in the implementation of MF-DFA \cite[][]{Kantelhardt:2002}.  Firstly, one constructs a nonstationary {\em profile} $Y(i)$ of the original time series $X_i$, which is the cumulative sum 
\begin{equation}
Y(i )\equiv \sum_{k=1}^{i} \left(X_k - \bar{X_k} \right), \qquad \text{where}\qquad  i = 1, ... , N.  
\label{eq:profile}
\end{equation}
Secondly, the profile is divided into $N_s = \text{int}(N/s)$ segments of equal length $s$ that do not overlap.  Excepting rare circumstances, the original time series is not an exact multiple of $s$ leaving excess segments of $Y(i)$.  These are dealt with by repeating the procedure from the end of the profile and returning to the beginning and hence creating $2 N_s$ segments.  Thirdly, within each of the $\nu = 1,... 2N_s$ segments the variance $\text{Var}(\nu, s)$ of the profile relative to a local least squares polynomial fit $y_{\nu}(i)$ of $n^\text{th}$ order is
\begin{equation}
\text{Var}(\nu, s) \equiv \frac{1}{s} \sum_{i=1}^{s} \left[ Y([\nu - 1]s + i) - y_{\nu}(i)\right]^2. 
\label{eq:var}
\end{equation}
Finally, the generalized fluctuation function is formed as
\begin{equation}
F_q (s) \equiv \left[ \frac{1}{2 N_s} \sum_{\nu=1}^{2 N_s} \{ \text{Var}(\nu, s)\}^{q/2} \right]^{1/q},  
\label{eq:fluct}
\end{equation}
and the principal tool of MF-DFA$n$ is to examine how $F_q (s)$ depends on the choice of time segment $s$ for a given degree of polynomial fit $n$ and the order $q$ of the moment taken.  The scaling of  the generalized fluctuation function is characterized by a generalized Hurst exponent $h(q)$ viz., 
\begin{equation}
F_q (s) \propto s^{h(q)} .  
\label{eq:power}
\end{equation}
When the time series is monofractal then $h(q)$ is independent of $q$ and is thus equivalent to the classical Hurst exponent $H$.  For the case of $q$ = 2, MF-DFA and DFA are equivalent \cite[][]{Kantelhardt:2002}.  Whence, a time series with long-term persistence has $h(2) = 1 - \gamma/2$ for $0 < \gamma < 1$.  However, short-term correlated data, decaying faster than $1/s$, has $\gamma > 1$ and finite ${\cal{T}}_s$ leading to a change at $s = s^\star$, and asymptotic behavior defined by $h(2) = 1/2$.  Moreover, the connection between $h(2)$ and the decay of the power spectrum $S(f) \propto f^{- \beta}$, with frequency $f$ is $h(2) = (1 + \beta)/2$ \cite[e.g.,][]{Ding}.  Therefore, one sees that for classical white noise, $\beta$ = 0 and hence $h(2) = 1/2$, whereas for Brownian or red noise $\beta$ = 2 and $h(2) = 3/2$. 

In general multifractal time series exhibit a scaling dependence on the moment $q$.  One can relate the generalized Hurst exponents that we focus on here to other multifractal exponents \cite[e.g.,][]{Jens} and this is the subject of a separate publication.  Such complimentary exponents are useful because (a) different exponents are more easily extracted from different data sets and (b) multifractality can originate both in a broad probability density as well as large and small fluctuations having a different long-term persistence.  The MF-DFA procedure can distinguish between these \cite[][]{Kantelhardt:2002} and thus different exponents can provide tests of distinct multifractal origins.   Here, in Section \ref{sec:results}\ref{sec:Hurst},  we focus on the generalized Hurst exponents, which make the most transparent contact with the observations that are the topic of our study.

\subsection{Moving Windows and MF-TWDFA \label{sec:new}}

The generalization of MF-DFA by \citet[][]{Zhou:2010} applies a variant of the weighted least squares approach to fitting the polynomial $y_{\nu}(i)$ to the profile $Y(i)$ on each interval $\nu$.   Here, rather than using $n$th order $y_{\nu}(i)$'s to estimate $Y(i)$ {\em within} a fixed window, without reference to points in the profile outside that window, a moving window which is smaller than $s$ but determined by distance between points is used to construct a point by point approximation $\hat{y}_{\nu}(i)$ to the profile.  Thus, instead of Eq. \ref{eq:var} we compute the variance up ($\nu = 1,...,N_s$) and down ($\nu = N_s + 1,...,2 N_s$) the profile as
\begin{equation}
\begin{array}{l l}
\text{Var}(\nu, s)  \equiv \frac{1}{s} \sum_{i=1}^{s} \{ Y([\nu - 1]s + i) - {\hat{y}}([\nu-1]s +i) \}^2\\
\\
\text{for $\nu = 1,...,N_s$, and}\\
\\
\text{Var}(\nu, s)  \equiv \frac{1}{s} \sum_{i=1}^{s} \{ Y(N-[\nu - N_s]s + i) - {\hat{y}}(N-[\nu-N_s]s +i)\}^2\\
\\
\text{for $\nu = N_s + 1,...,2 N_s$.}\\
\end{array}
\label{eq:varTW}
\end{equation}
Therefore we replace the global linear regression of fitting the polynomial $y_{\nu}(i)$ to the data as appears in Eq. (\ref{eq:var}), with a weighted local estimate $\hat{y}_{\nu}(i)$ determined by the proximity of points $j$ to the point $i$ in the time series such that $\vert i - j \vert \le s$.   A larger (or smaller) weight $w_{ij}$ is given to $\hat{y}_{\nu}(i)$ according to whether $\vert i - j \vert $ is small (large). 

In the ordinary least squares method one minimizes the sum of the squared differences between the predicted and the original function, whereas in weighted least squares a weight factor is applied to the squared differences before performing the minimization.  There are many varieties of weighted least squares originating in time series analysis for evenly spaced observations \cite[][]{Macauley:1931} and ``robust'' locally weighted regression for unevenly spaced data \cite[][]{Cleveland:1979}.   The standard weighted linear regression scheme fits observed variables $Y(i)$ and ${{\cal X}_k}(i)$ assuming a linear relationship as
\begin{equation}
Y(i) = a_0(i) + \sum_{k=1}^{m} {a_k}(i){{\cal X}_k}(i) + \epsilon(i), 
\end{equation}
where $\epsilon(i)$ is the error, which may (but need not) be homoscedastic.  Here,  ${a_k}(i)$ is the $k^\text{th}$ fitting parameter at time $i$, the vector $\bf a$ of which is determined from 
\begin{equation}
\bf{a} = ({\bf {\cal X}}^\text{T} {\bf w} {\bf {\cal X}})^{-1} {\bf {\cal X}}^\text{T} {\bf w} {\bf Y}, 
\label{eq:coeffs}
\end{equation}
where $\bf w$ is a diagonal matrix with weight elements $w_{ij}$ with magnitudes that depend on the proximity to $i$ thereby estimating an ${a_k}(i)$ according to a prescribed criterion of proximity. 
Hence, $\hat{y}_{\nu}(i)$ is determined by a weighted least squares approach using
\begin{equation}
\hat{y}_{\nu}(i) = a_0(i) + a_1(i) i + \epsilon(i),  \qquad i = 1, ... , N. 
\end{equation}
Here again the fitting parameters $a(i) = [a_0(i), a_1(i)]^\text{T}$ depend on the place $i$ in the series and are determined from equation (\ref{eq:coeffs})
in which the $w_{ij}$, with $j = 1, ... , n$, are the diagonals of an $n \times n$ diagonal matrix, and $\bf{\cal X}$ is an $n \times 2$ matrix with the first column being unity and the second running from 1 to $n$.   We use the same bisquare function for the weights
\begin{equation}
w_{ij} =
\begin{cases}
\left( 1 - \left[\frac{i - j}{s}\right]^2\right)^2, & \textrm{if}~ \vert i - j \vert \le s,\\
0, & \textrm{otherwise,}
\end{cases},
\label{eq:bisquare}
\end{equation}
as did \citet[][]{Zhou:2010}, although one can envisage a range of possibilities for dealing with the temporal weighting. 

We conclude this section by emphasizing that a principal advantage of MF-TWDFA over conventional MF-DFA is the robustness of extracting the scaling of the fluctuation function, and hence the crossover points between scalings, that indicate the underlying processes reflected in the data.  Of particular importance is the fidelity of extracting the long-term scaling behavior in geophysical data sets which can often be obscured in MF-DFA and, as we shall see below, compromises MF-TWDFA when a strong seasonal cycle remains in the original time series.  Moreover, in MF-DFA the profile of the time series is fit using discontinuous polynomials, which can introduce errors in the determination of crossover times for new scalings, and can be particularly questionable at long time scales.  Finally, for time series of length $N$, whilst MF-DFA is typically informative only up to $N/4$, MF-TWDFA can be carried out to $N/2$. 

\section{Sea Ice Geophysical Data \label{sec:data}}

\begin{figure}
\centering
\includegraphics[width=8.5cm]{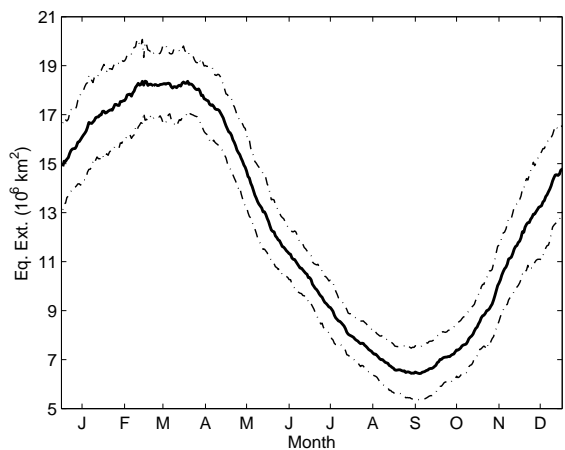}
\caption{The mean seasonal cycle of the Equivalent Ice Extent (EIE) during the satellite era (Fig. \ref{fig:coverage}) shown as the solid line, with the dashed lines denoting one standard deviation.}
\label{fig:seasonal}
\end{figure}

\begin{figure}[h!]
\centering
\includegraphics[width=8.5cm]{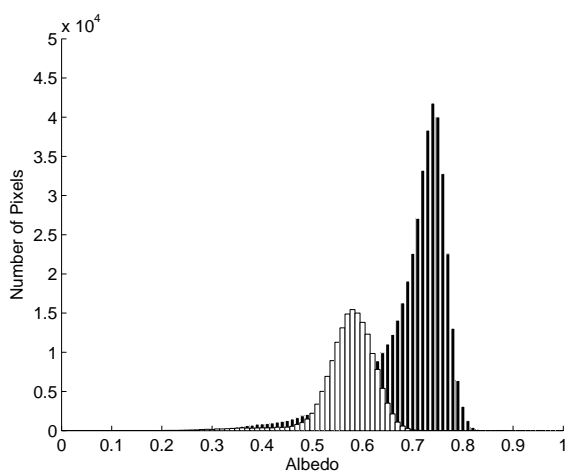}
\caption{Albedo histograms shown for days in mid-March (black) and -September (white).  If there is ice in a pixel for the 23 year record then we compute the albedo for that pixel.  We compute a histogram for each day of the year from the number of pixels for each average albedo bin for that day \cite[][]{Sahil:grl}.}
\label{fig:hist}
\end{figure}

We use MF-TWDFA to examine the multi-scale structure of two satellite based geophysical data sets for Arctic sea ice;  the Equivalent Ice Extent (EIE) and albedo retrievals from the Advanced Very High Resolution Radiometer (AVHRR) Polar Pathfinder (APP) archive.  The EIE data derives from retrievals of satellite passive microwave radiances over the Arctic converted to daily sea ice concentration using the NASA Team Sea Ice Algorithm in the same manner as described by \citet[][]{IanGeom}, who focused on the origin of the difference between ice extent and EIE.   We refer the reader to Eisenman's paper for a detailed description of the determination of EIE.  The mean EIE seasonal cycle from 1978-present is shown in Fig. \ref{fig:seasonal}.  Daily satellite retrievals of the directional - hemispheric apparent albedo are determined from the APP archive as described in a separate publication dedicated to a different form of analysis than presented here \cite[][]{Sahil:grl}.  The apparent albedo is what would be measured by upward and downward looking radiometers and thus varies with the state of the atmosphere and the solar zenith angle.

\begin{figure}

      \begin{center}
	  \subfigure(a){%
	      \includegraphics[width=5.7cm,height=3.7cm]{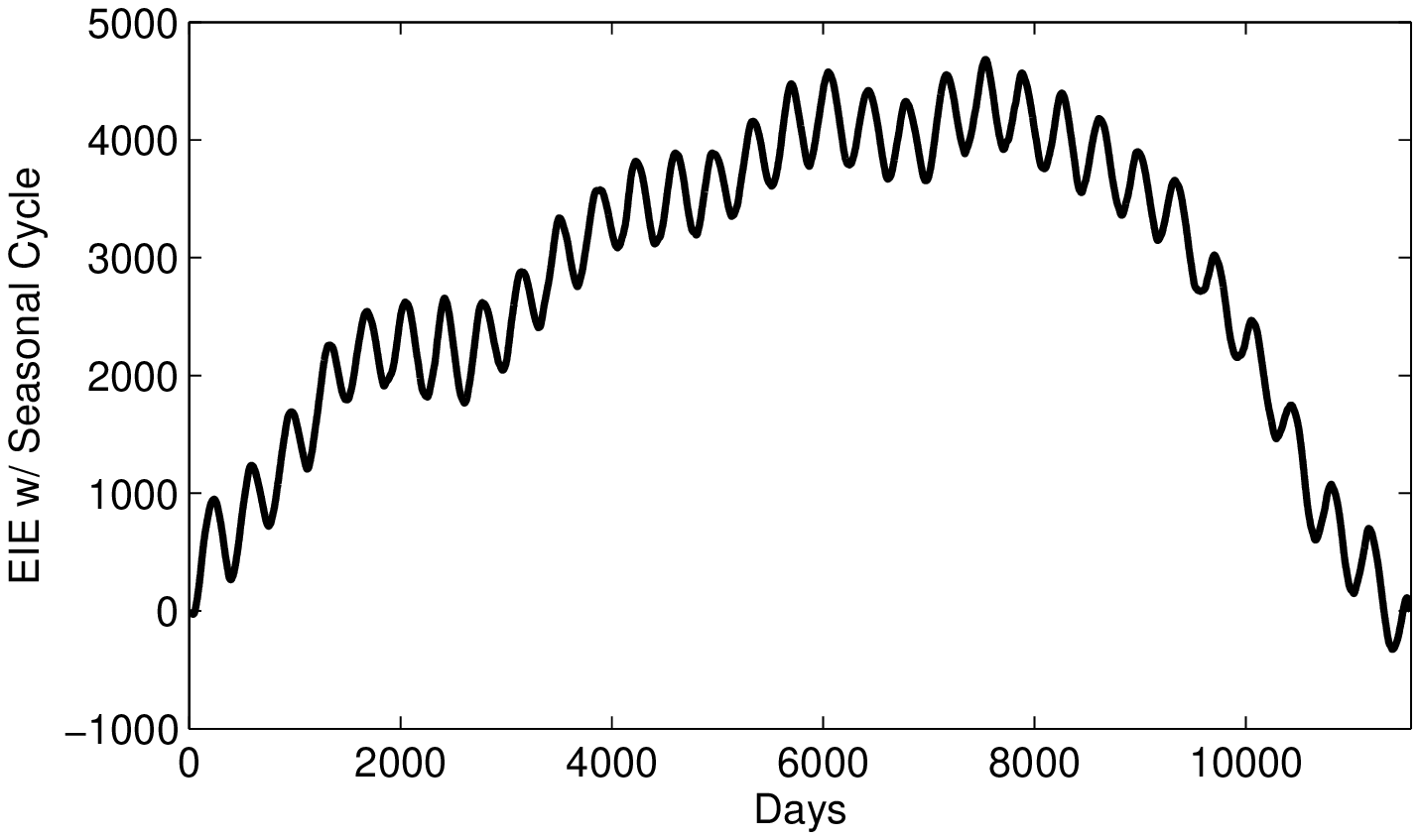}
	  }
	  \subfigure(b){%
	    \includegraphics[width=5.7cm,height=3.7cm]{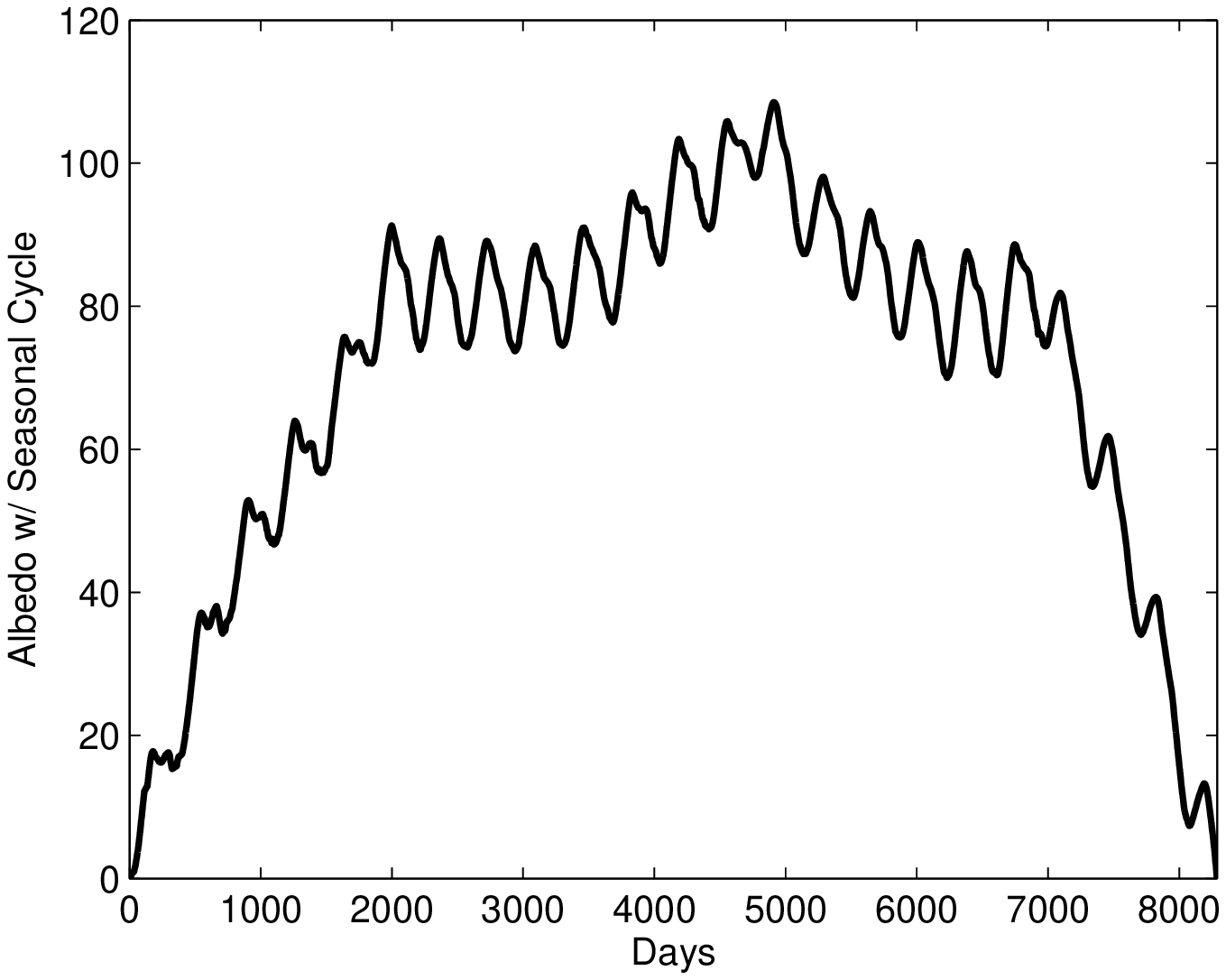}
	  }\\ 
 	  \subfigure(c){%
	      \includegraphics[width=5.7cm,height=3.7cm]{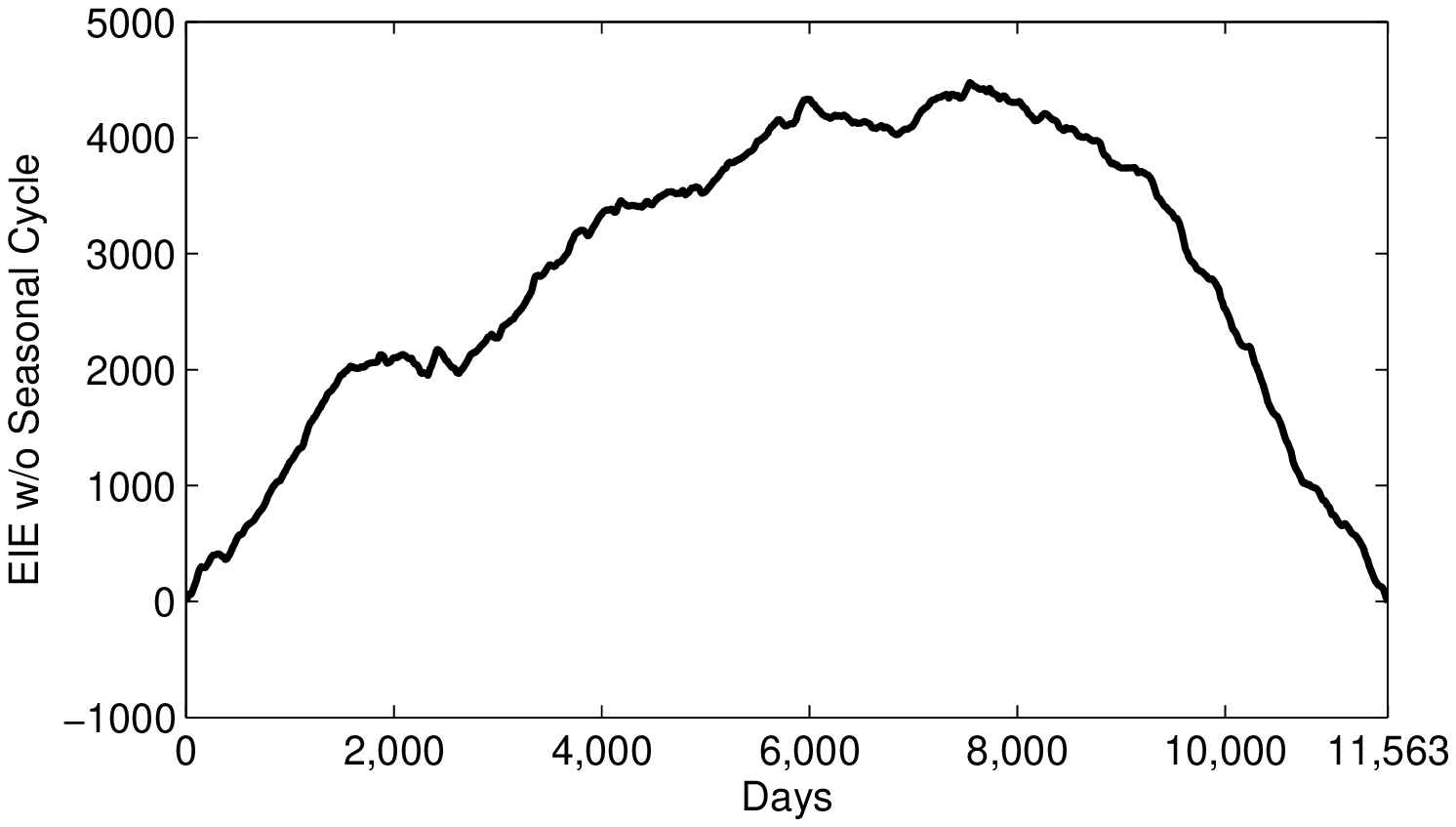}
	  }
	  \subfigure(d){%
	    \includegraphics[width=5.7cm,height=3.7cm]{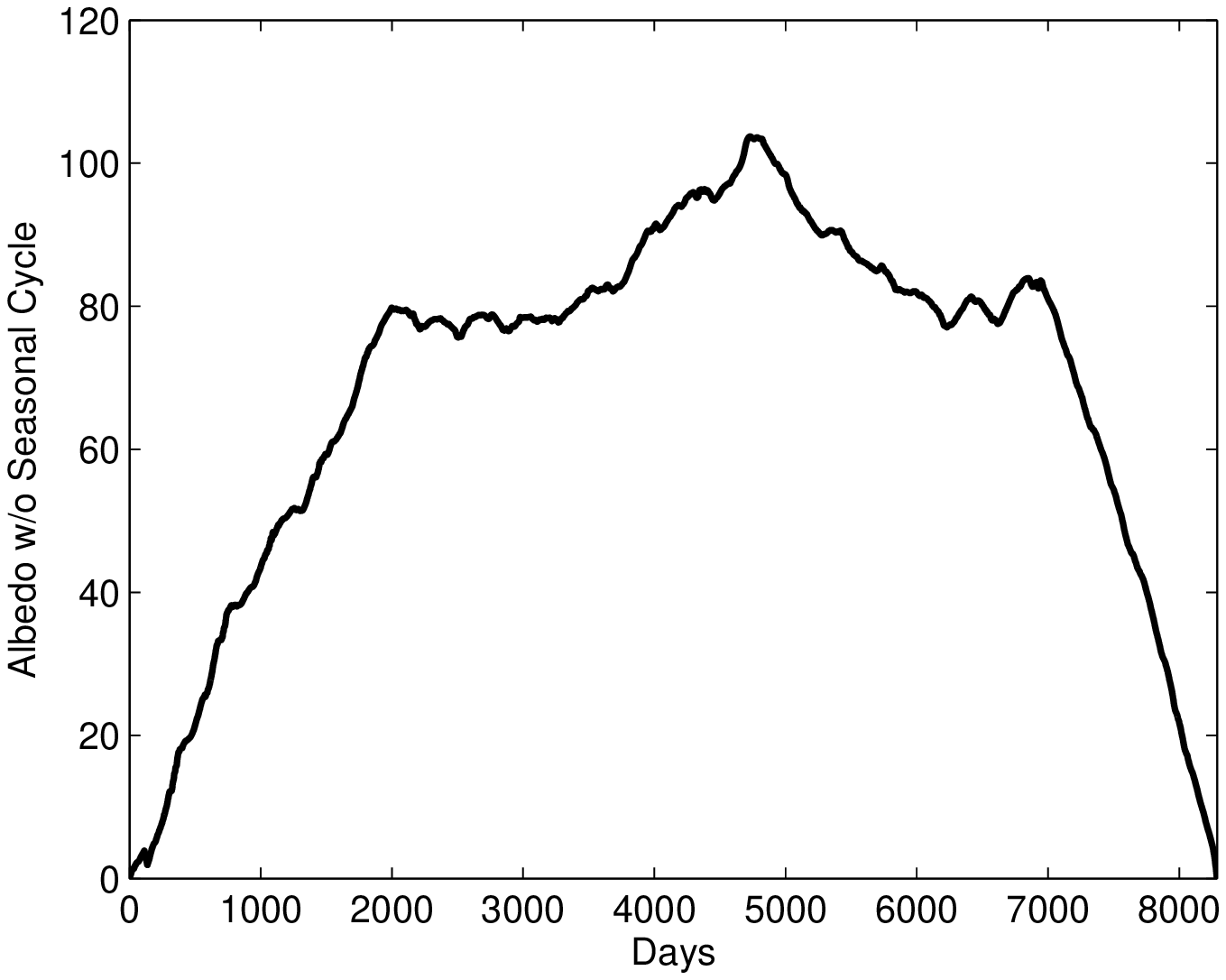}
	  }\\ 
      \end{center}
      \caption{(a) The profile for the EIE  and (b) albedo using Eq. \ref{eq:profile} with a seasonal cycle in the original time series.  (c) The profile for the EIE and (d) albedo using Eq. \ref{eq:profile} {\em removing} the seasonal cycle from the original time series. 
      }%
       \label{fig:Profiles}
\end{figure}

The APP dataset has been refined for use in a wide range of polar studies and is described in detail in \citet[][and refs. therein]{Sahil:grl}.  In brief, the AVHRR channels range from the visible to the thermal infrared (0.58 - 12.5 $\mu$m) and measure top of the atmosphere reflectances  and brightness temperatures.  With 5~km $\times$ 5~km resolution, we analyze albedo retrievals from 1 January 1982 through 31 December 2004, taken daily at 1400 hours.  Sea ice is distinguished from land and open water using microwave brightness temperatures and filtering the Surface Type Mask data with the NASA Team Sea Ice Algorithm which distinguishes between first-year (FYI) and multi-year  ice (MYI) concentrations.  The approach ascribes a MYI flag to a region containing at least 50 percent of this ice type, with uncertainties depending on the season (e.g., melt pond fraction) and region (near the ice edge), along with the surface type categories.  On physical grounds the albedo data are filtered to remove any values greater that 1 or less than 0.2. Each pixel is assessed every day for the presence of ice and then albedo is averaged for that pixel \cite[][]{Sahil:grl}. Examples of histograms for mid-March and mid-September are shown in Fig. \ref{fig:hist}.

\section{Results and Discussion \label{sec:results}}

Here we focus most of our discussion on the nature of the fluctuation functions $F_q (s)$ that emerge from MF-TWDFA and return later to the exponents, generalized dimensions and singularity spectra.  
First, we note  that we have eliminated the possibility of spurious multi-fractility \cite[][]{Schumann:2011} by checking the usual measures $\Delta h_{20} \equiv h(- 20)-h(20)$ and $\Delta\alpha$.  Second, whilst we show the fluctuation functions for a range of $q$, for clarity we also show plots solely for $q=2$ and remind the reader that for  $h(2) = 1/2$ the system exhibits a completely uncorrelated white noise dynamics, and for $h(2) = 3/2$ there are red (or Brownian) noise correlations.  Whereas for  $0 < h(2) < 1/2$ the dynamics are anticorrelated.  Moreover, as noted above, for analysis by power spectra $\beta = 2 h(2) -1$.   It is evident from equation (\ref{eq:fluct}) that small temporal fluctuations are characterized by $h(q)$'s for $q<0$ whereas large temporal fluctuations are characterized by $h(q)$'s for $q>0$.  As we discuss below the fluctuation functions $F_q (s)$ do indeed  distinguish the scaling of small and large fluctuations.  

\begin{figure}
      \begin{center}
	  \subfigure(a){%
	      \includegraphics[width=.45\textwidth]{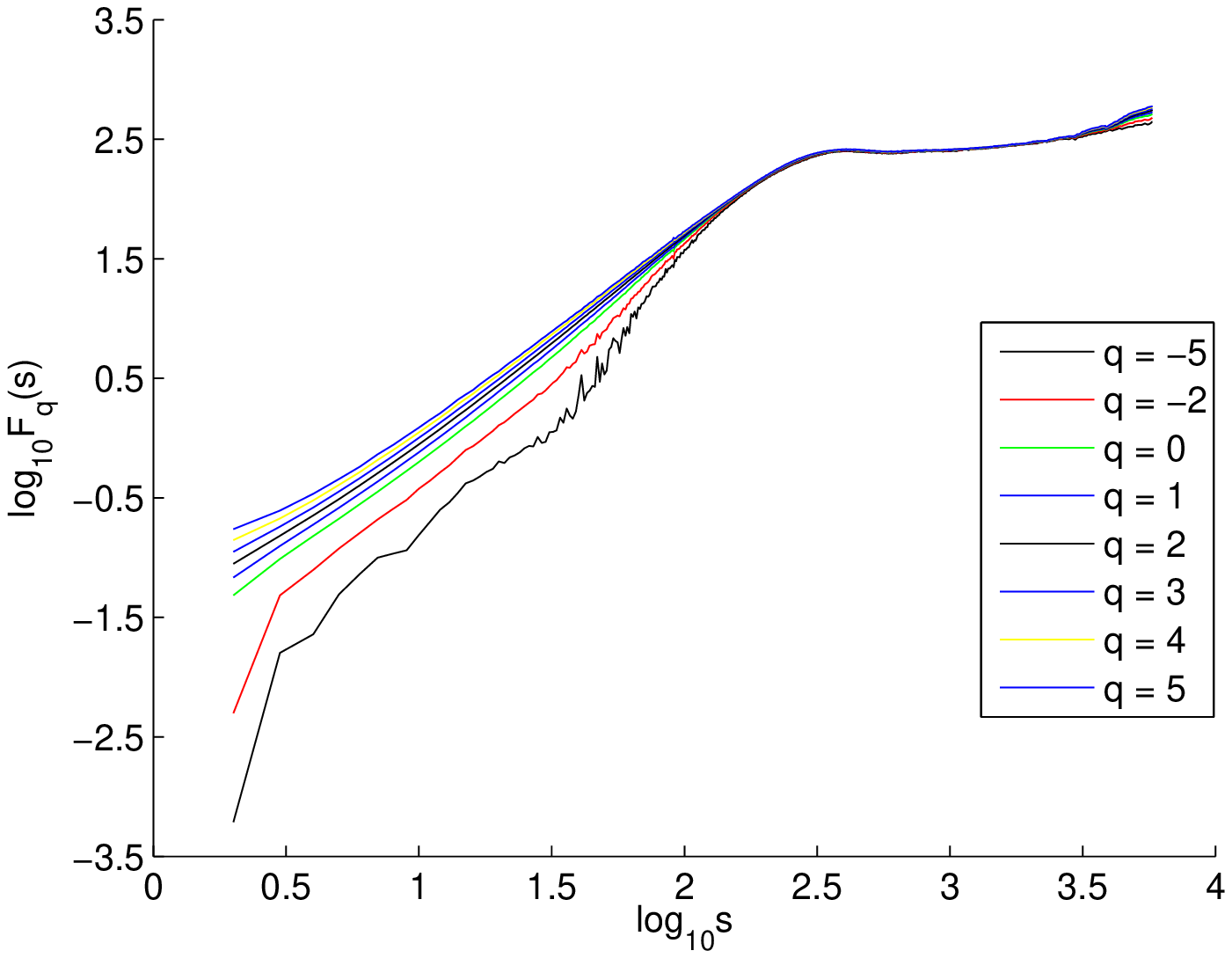}
	  }
	  \subfigure(b){%
	    \includegraphics[width=.45\textwidth]{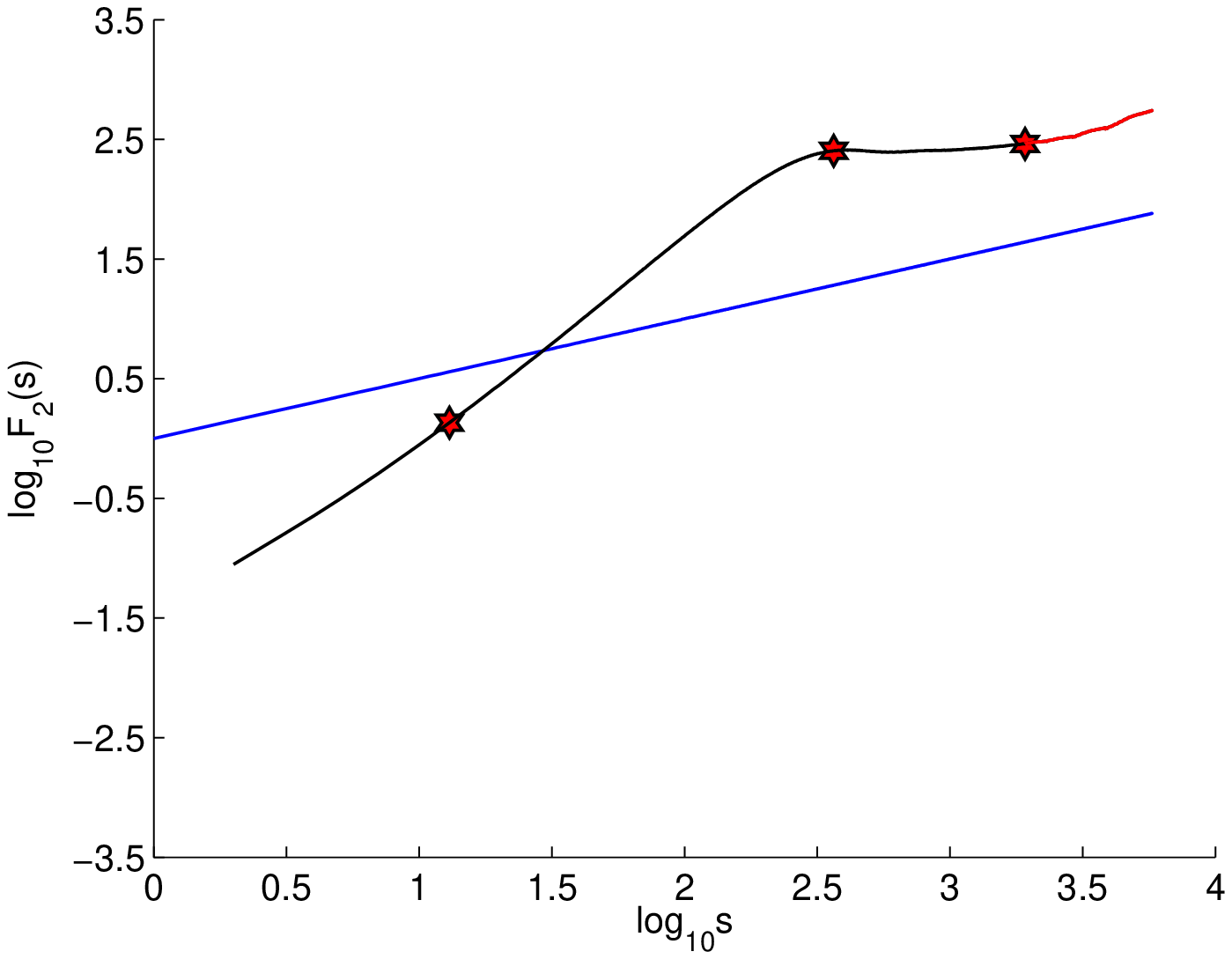}
	  }\\ 
	  	  \subfigure(c){%
	      \includegraphics[width=.45\textwidth]{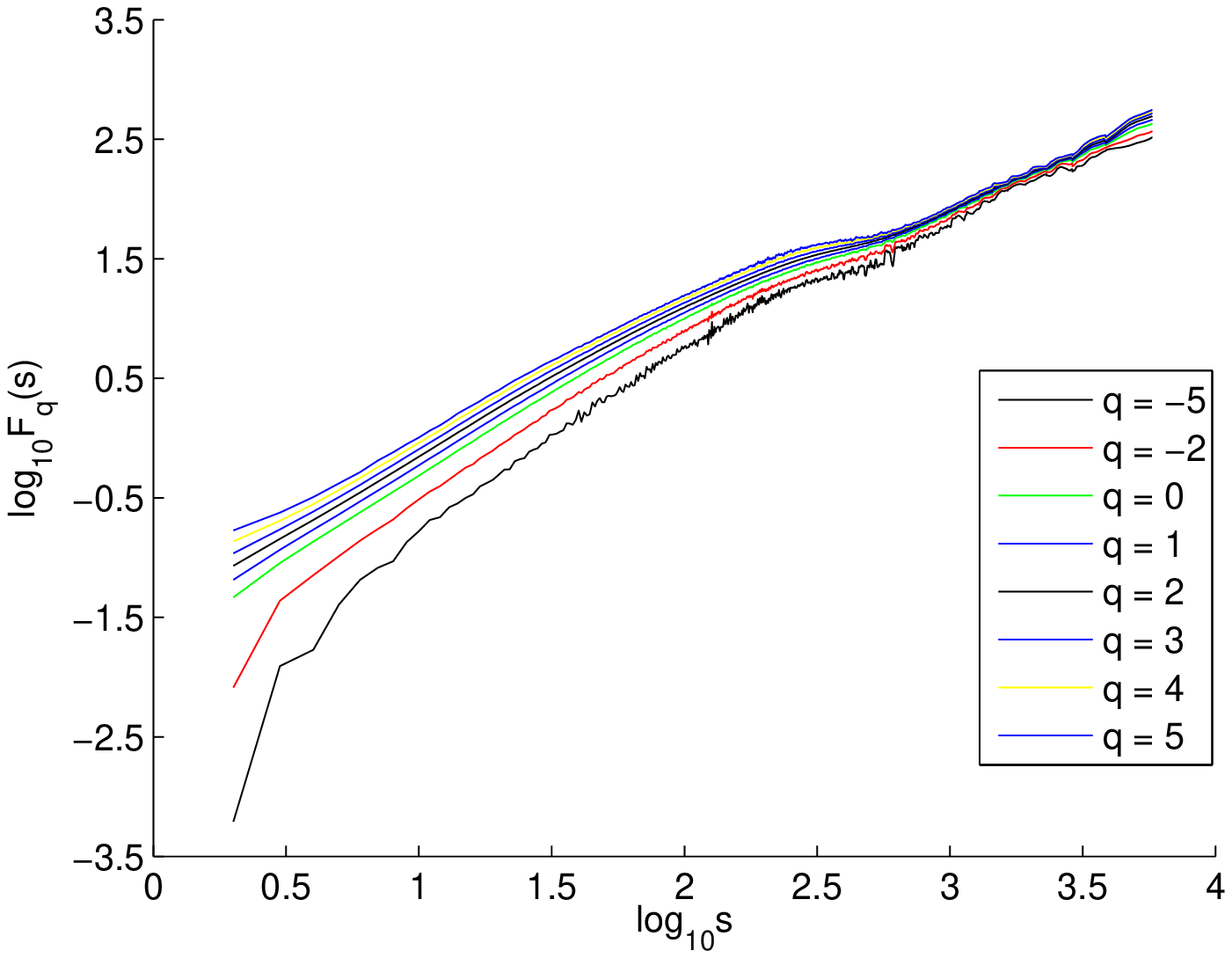}
	  }
	  \subfigure(d){%
	    \includegraphics[width=.45\textwidth]{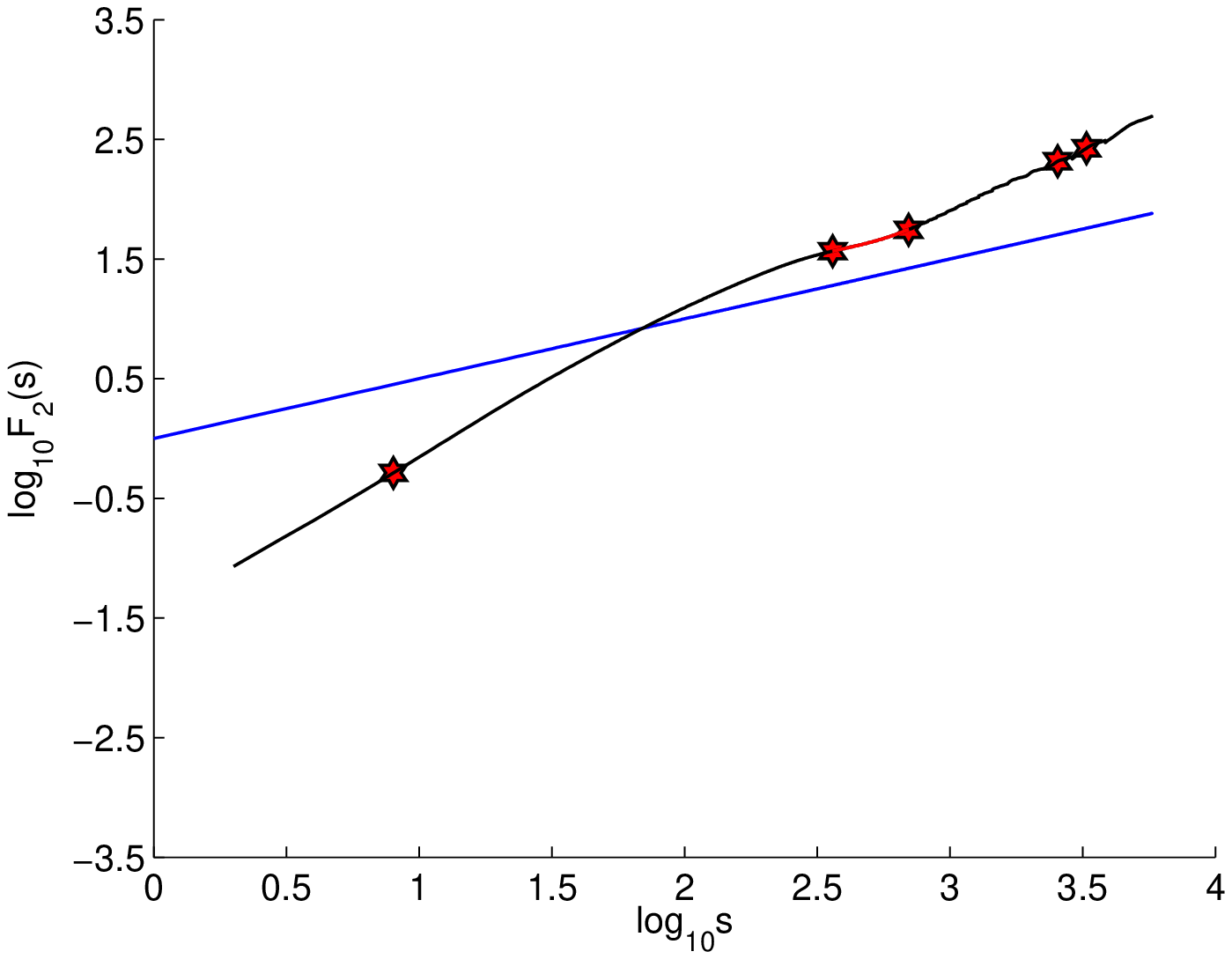}
	  }\\ 
      \end{center}
      \caption{(a). The fluctuation function from equation (\ref{eq:power}) for the Equivalent Ice Extent (EIE) {\em with a seasonal cycle}.  The $q$'s are shown in the panel and $s$ is measured in days throughout. (b). The fluctuation function from equation (\ref{eq:power}) for Equivalent Ice Extent (EIE) {\em with a seasonal cycle} for $q$=2.  The stars denote the crossover times associated with a slope change at approximately 13 days, 1 year and 5.25 years.  The blue line and the red segment of the black curve both denote white noise with $h(2)$ = 1/2.  (c) and (d) correspond to (a) and (b) but {\em without a seasonal cycle}.  Here, the stars denote crossover times of approximately 8 days,  1 year,  2 years,  6.9 years and 8.9 years.  Again, the blue line and the red segment of the black curve both denote white noise with $h(2)$ = 1/2.
      }%
      \label{fig:EIEwss}
\end{figure}
\begin{figure}

      \begin{center}
	  \subfigure(a){%
	      \includegraphics[width=.45\textwidth]{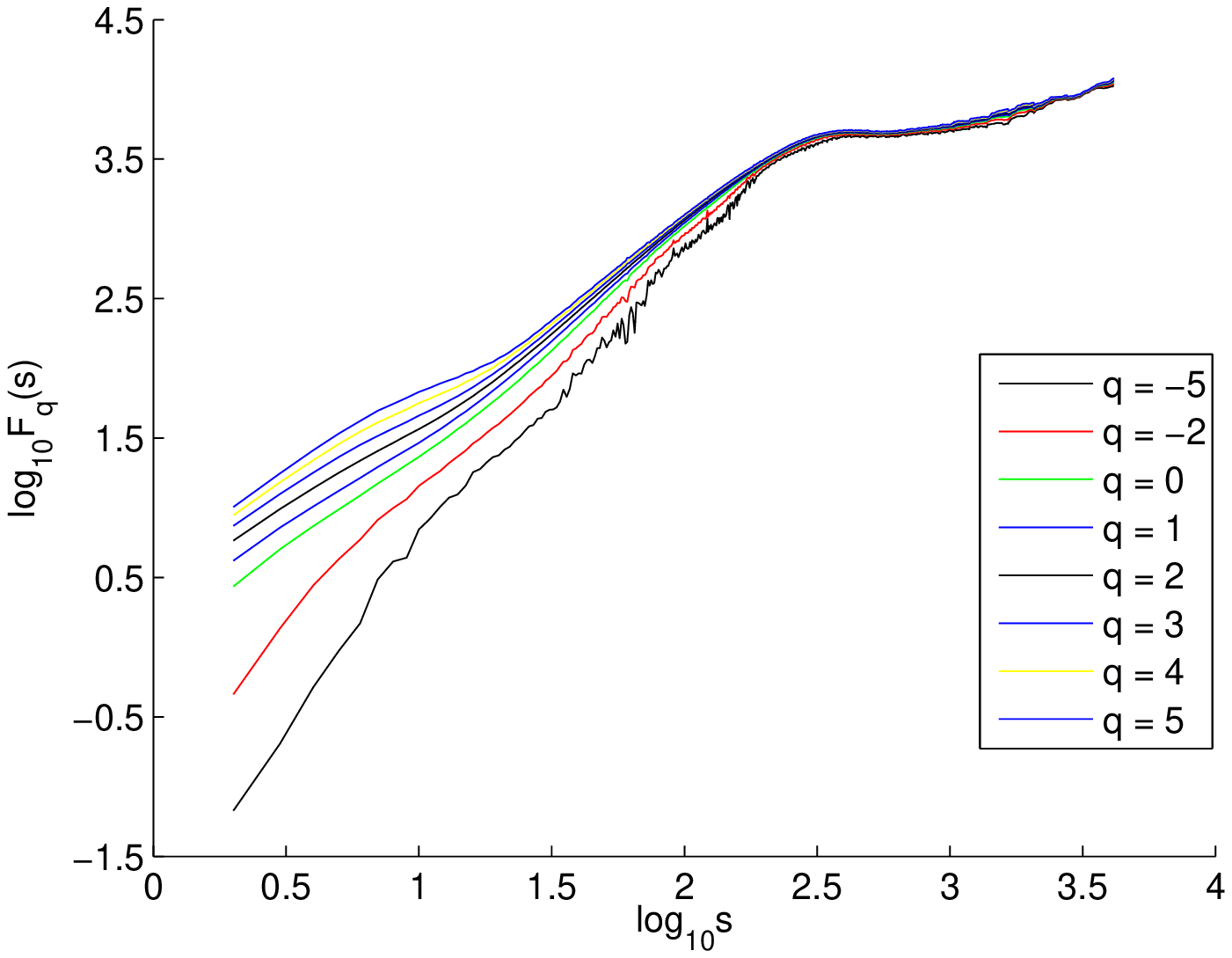}
	  }
	  \subfigure(b){%
	    \includegraphics[width=.45\textwidth]{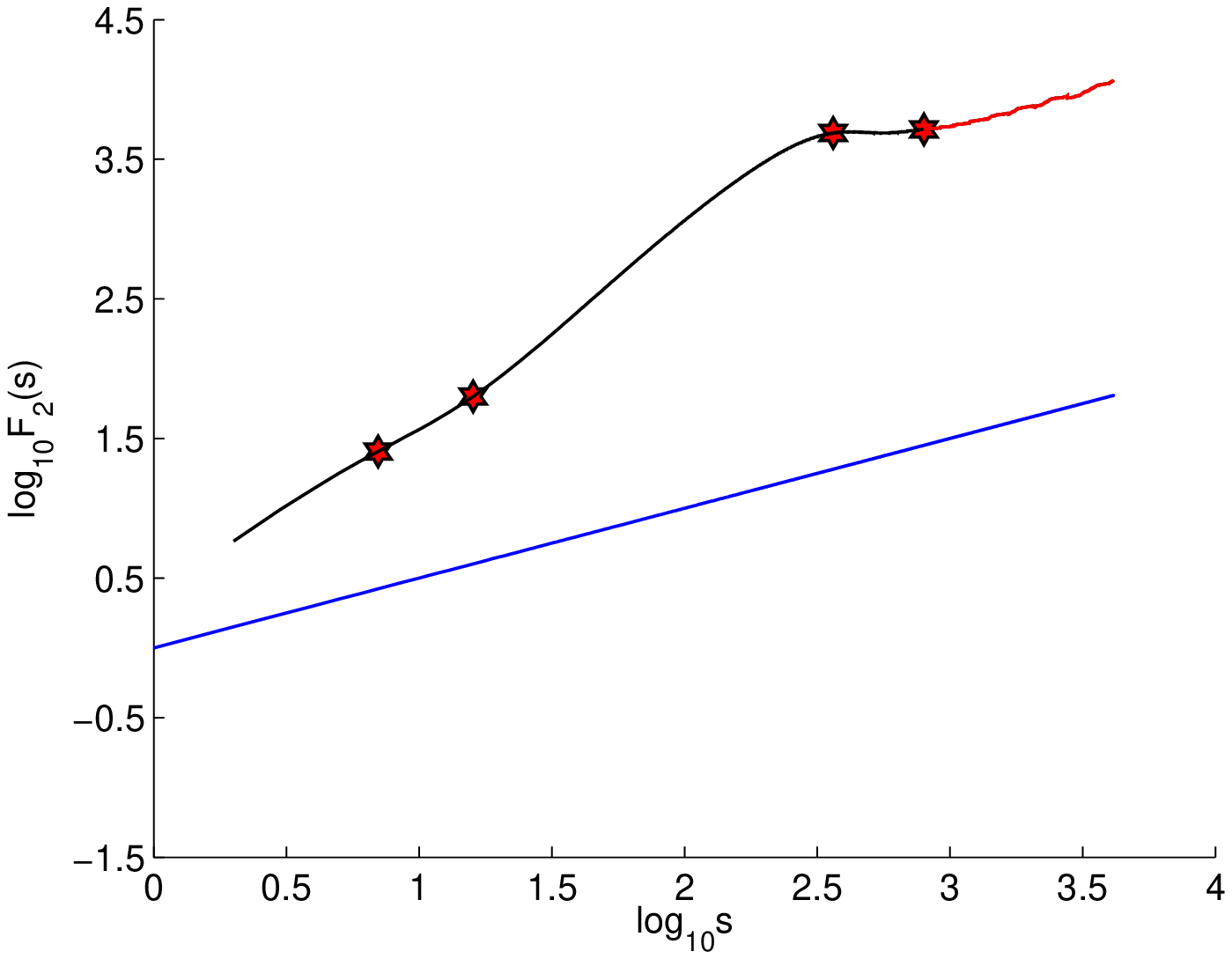}
	  }\\ 
	  	  \subfigure(c){%
	      \includegraphics[width=.45\textwidth]{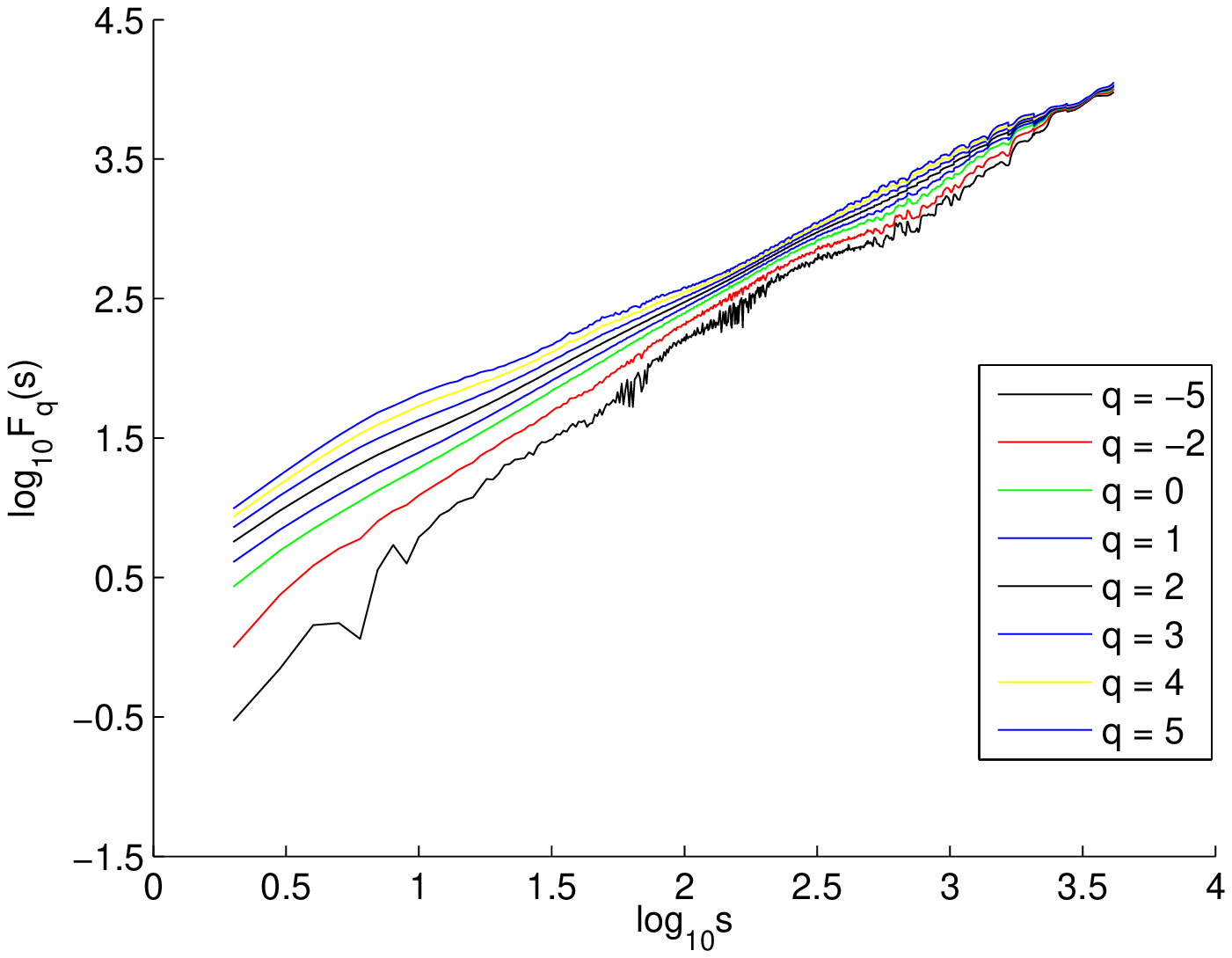}
	  }
	  \subfigure(d){%
	    \includegraphics[width=.45\textwidth]{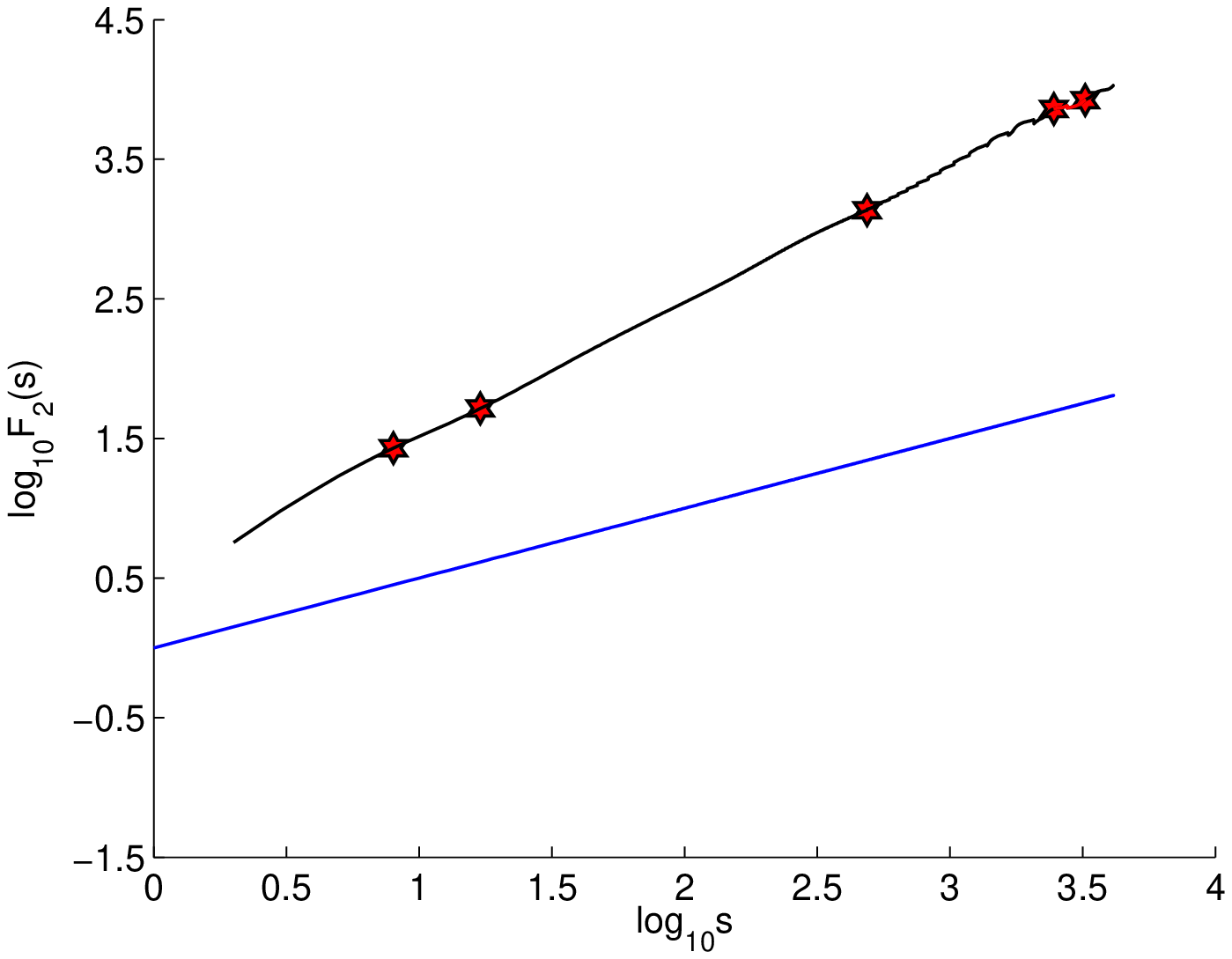}
	  }\\ 

      \end{center}
      \caption{(a). The fluctuation function from equation (\ref{eq:power}) for the ice albedo {\em with a seasonal cycle}.  The $q$'s are shown in the panel.  (b). The fluctuation function from equation (\ref{eq:power}) for ice albedo {\em with a seasonal cycle} for $q$=2.  The stars denote the crossover times associated with a slope change at approximately 7 days,  16 days,  1 year and 2.2 years. The blue line and the red segment of the black curve both denote white noise with $h(2)$ = 1/2.  (c) and (d) correspond to (a) and (b) but {\em without a seasonal cycle}.   The stars denote crossover times of approximately 8 days,  17 days,  487 days,  7.5 years and 9 years.  Here again, the blue line and the red segment of the black curve both denote white noise with $h(2)$ = 1/2.  
      }%
      \label{fig:albdwss}
\end{figure}

On all segments of the $F_q (s)$ plots the slopes are computed and the ``crossovers'' refer to the times $s$ where the curve changes slope.  As a test of the fidelity of the many crossovers in slope (associated with particular instrinsic time scales) detected with MF-TWDFA, we applied {\em both} MF-DFA and MF-TWDFA.  We found that  the crossovers for time scales of  2 years or longer would not have been captured by  MF-DFA because of large amplitude fluctuations in this range.  Note, for the EIE data with a seasonal cycle this result was confirmed for polynomial fitting in the detrending step of MF-DFA of up to order 9, while for the other data up to order 3.  Moreover, and that which we focus on here, the MF-TWDFA analysis used on the time series after removing the seasonal cycle leads to the extraction of crossovers associated with long term persistence that are ``masked'' when the seasonal cycle is not removed.  Next we discuss this finding in more detail. 

\subsection{Masking long term correlations by strong seasonal cycle  \label{sec:masking}}

A prominent feature of our analysis is the role played by the seasonal cycle.  We lay bare the distinction between profiles with and without the seasonal cycle in Figs. \ref{fig:Profiles}, which is the origin of the striking distinction between the upper and lower panels of Figs. \ref{fig:EIEwss} and \ref{fig:albdwss} for the EIE and albedo respectively.   It is seen that the strength of the seasonal cycle is such that it ``masks'' the dynamics on time scales longer than seasonal thereby suppressing the fingerprints of long term persistence.  This is clearly displayed in the fluctuation functions.  Firstly, 
in Figs. \ref{fig:EIEwss} and \ref{fig:albdwss} panels (a) and (b) which have a seasonal cycle, we see that the curves for all $q$'s converge at the 1 year time scale.  This convergence is removed when the seasonal cycle is removed as seen in Figs.  \ref{fig:EIEwss} and  \ref{fig:albdwss} panels (c) and (d). Secondly, when the original time series still possesses the seasonal cycle,  the slope abruptly drops below $h(2)$ = 1/2 at the 1 year time scale to transition to an anticorrelated structure.  This behavior is distinct from the clear transitions and positive slopes at longer time scales that are seen when the seasonal cycle is removed from the original time series.  Whilst there is a finite union of underlying processes that influence the ice extent and the ice albedo, that union is clearly not one to one \cite[see discussion in][]{Sahil:grl}.  Thus, although we find similar qualitative structure in the analysis of the EIE and albedo data with and without a seasonal cycle, there are important quantitative distinctions.  For example, the high frequency weather time scales of order week's are clearly seen in all of the data.  However, Figs. \ref{fig:EIEwss} and \ref{fig:albdwss} (a) and (b) clearly show masking beginning at one year, the transition to anticorrelated behavior and then a steeper--white noise--slope of $h(2)$ = 1/2 returning at 5.25 and 2.2 years in the EIE and albedo respectively.  Whence, all long term persistence is destroyed by the strength of the seasonal cycle which is associated with the strong periodicity remaining in the profiles of Fig. \ref{fig:Profiles}.   We see that the general steep/flat/steep behavior of the slopes of the fluctuation function with increasing time is seen in all panels of Figs. \ref{fig:EIEwss} and \ref{fig:albdwss}.   However, the essential distinction is that when the seasonal cycle is removed long term persistence is reentrant.  Indeed, time scales of $\sim$ 7 and 9 years are revealed in {\em both} the EIE and albedo data.  

\subsection{White noise on seasonal time scales  \label{sec:white}}

The region between 1 and 2 years in the EIE data without a seasonal cycle shows $h(2) \approx $ 1/2.  In other words, although the strength of the seasonal cycle is such that it dominates the power spectrum (not shown), when removing the seasonal scale from the original EIE data the system exhibits a white noise behavior from the seasonal to the bi-seasonal time scales.  However, the clear fingerprints of the short (weather) and long ($\sim$ 7 and 9 year) time scales remain.  The implications of this finding are, unfortunately for the goal of forecasting, rather clear;  a given ice minimum (maximum) can be followed by a minimum (maximum) with a larger or smaller magnitude. The same logic follows for minima/maxima separated by two years and the EIE states on time sca les between.   Most of the discussion in the literature focuses on the extremes in the observations (maxima and minima) which we find here to exhibit a nearly white time distribution.  If we interpret the results in the limit $h(2) = $ 1/2 then the autocorrelation will have strong peaks at the seasonal maxima, minima and bi-seasonal maxima and minima with rapid decays for times from one to two years.  As discussed below, this is consistent with aspects of other studies in which rapid decorrelations are found.  

It is important to be clear regarding what these results {\em do not} tell us.  We are not saying that existence of minima or maxima (or states between) are uncorrelated in time, but solely that the magnitudes of those states cannot be predicted to be larger or smaller from these data alone.  Moreover, the lack of an extended autocorrelation does not a-priori constrain the probability distribution(s) associated with the process(es) from which the value of the observable originates, but rather solely refers to the temporal distribution of the observable.  In other words whilst a particular probability distribution may underly the nature of how a
given magnitude of the EIE or albedo is reached, we find that the temporal distribution of the sequence of observed magnitudes is white on seasonal to bi-seasonal time scales. Thus, on such time scales these data are  {\em short-term correlated}, are associated with a finite ${\cal{T}}_s$ and an autocorrelation that decays faster than $\sim s^{- 1}$.  Therefore, on these short time scales, the processes controlling the EIE or albedo states in the Arctic are the familiar regular strong seasonal scale radiative and advective forcings.  Indeed, this is explained by the robustness of the response times associated with the seasonal control of the ice state through the competition between the ice-albedo feedback in summer and the long-wave forcing in winter \cite[e.g.,][]{MW:2011, Tietsche:2011}.  However, on longer time scales, we can associate the retreat of the ice cover with the observation of the upswing in $h(2)$ towards a more ``Brownian'' dynamics; $h(2)\sim3/2$.   Such trends, when accumulating future data over decadal time scales could well be compromised by a higher degree of noise in the state of the system, but to speculate more moves beyond the level of firm evidence. 

We test the idea that the rapid decline in the ice coverage during the most recent decade is dominating the longest crossovers found in our analysis as follows. The EIE record is approximately a decade longer than that for the albedo and covers the decade of the most rapid decline.  Thus we ask what crossovers are seen when we reanalyze this record between 1982-2004, the time range corresponding to the albedo data set?   It is found that the two long term crossovers decrease from 7 to 6.2 and from 9 to 8.4 years.  Moreover, the seasonal to bi-seasonality of the white noise persists.  Therefore, the increase in the crossover times as we include the recent data appears to be the most  plausible reflection of the recent declines.  However, we leave for future study what would likely be a less quantitative endeavor of correlating these longer time scales with climate indices that have time scales ranging from 2 to 7 years (for the El Ni\~{n}o Oscillation) to a decade or many decades (such as for the Pacific Decadal,  North Atlantic or Arctic Oscillations).

\begin{figure}

      \begin{center}
	  \subfigure(a){%
	      \includegraphics[width=.45\textwidth]{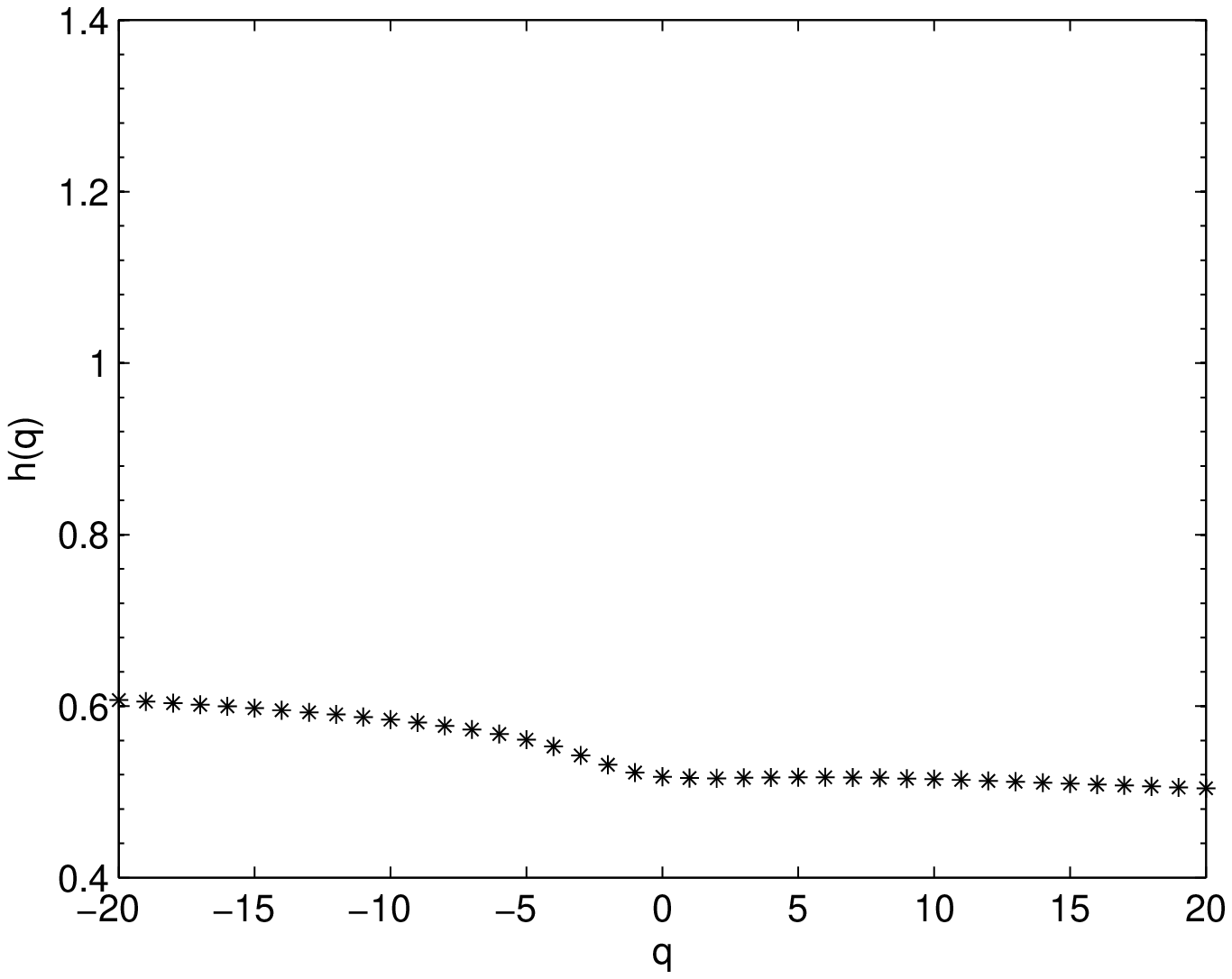}
	  }
	  \subfigure(b){%
	    \includegraphics[width=.45\textwidth]{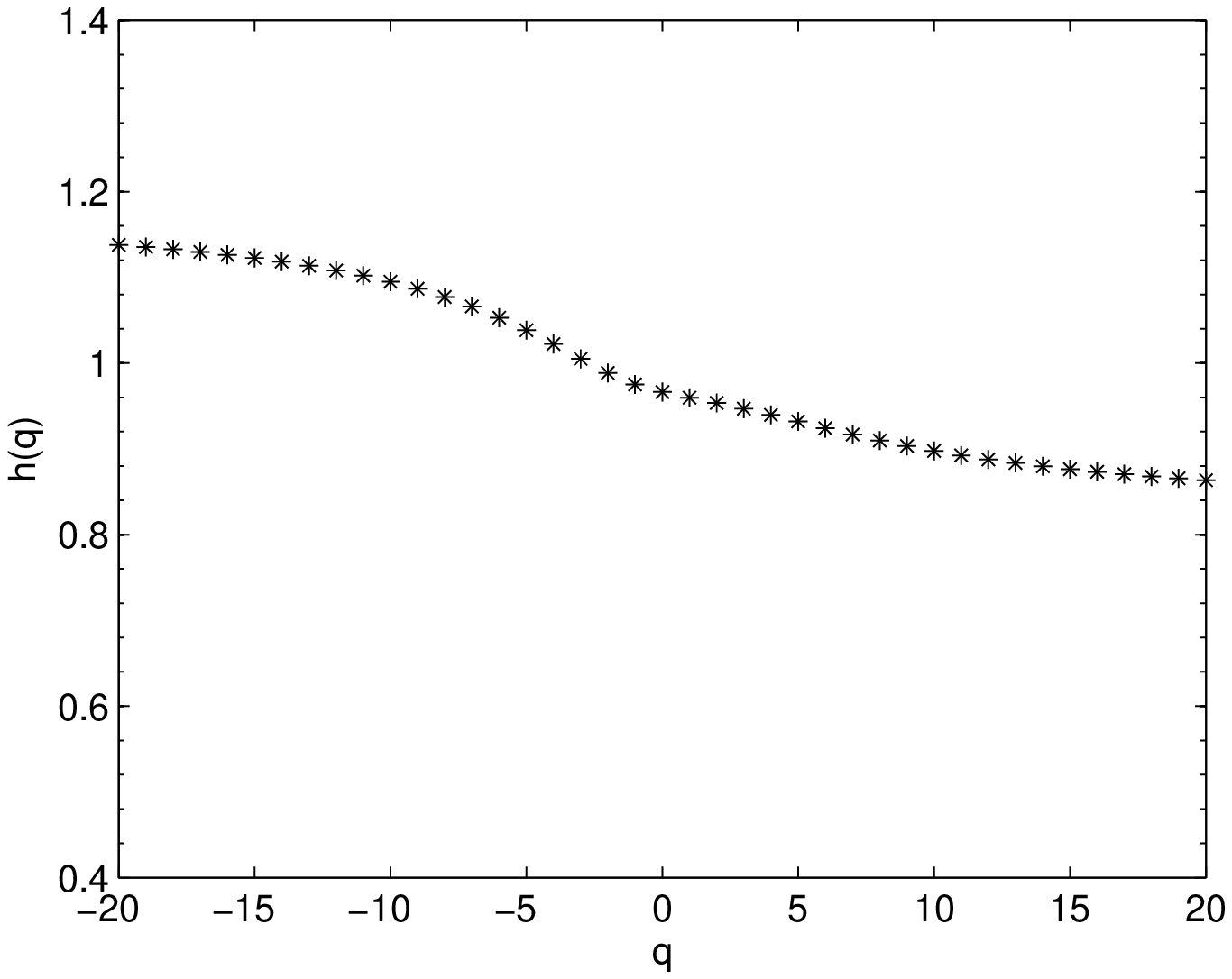}
	  }\\ 
	  \subfigure(c){%
	      \includegraphics[width=.45\textwidth]{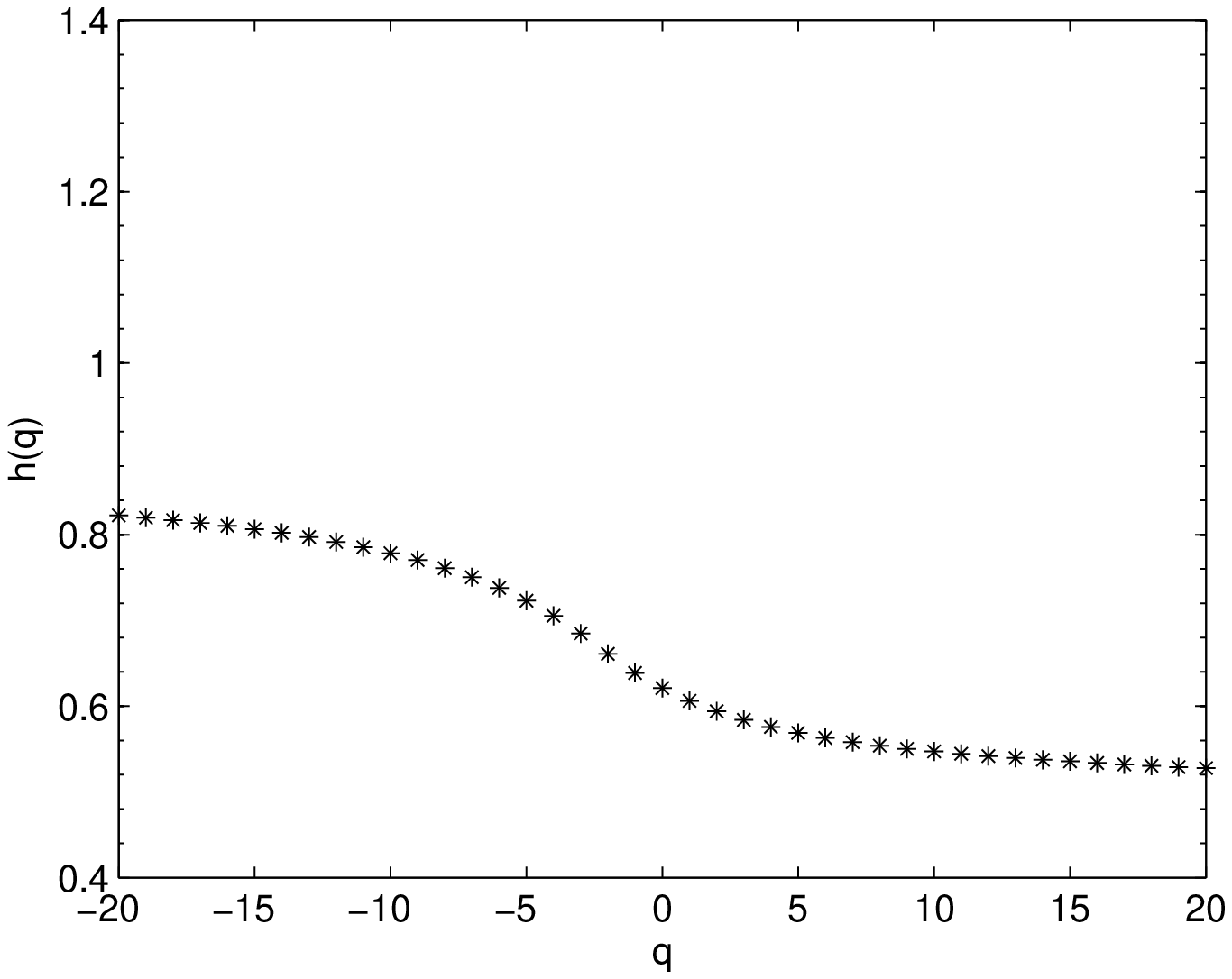}
	  }
	  \subfigure(d){%
	    \includegraphics[width=.45\textwidth]{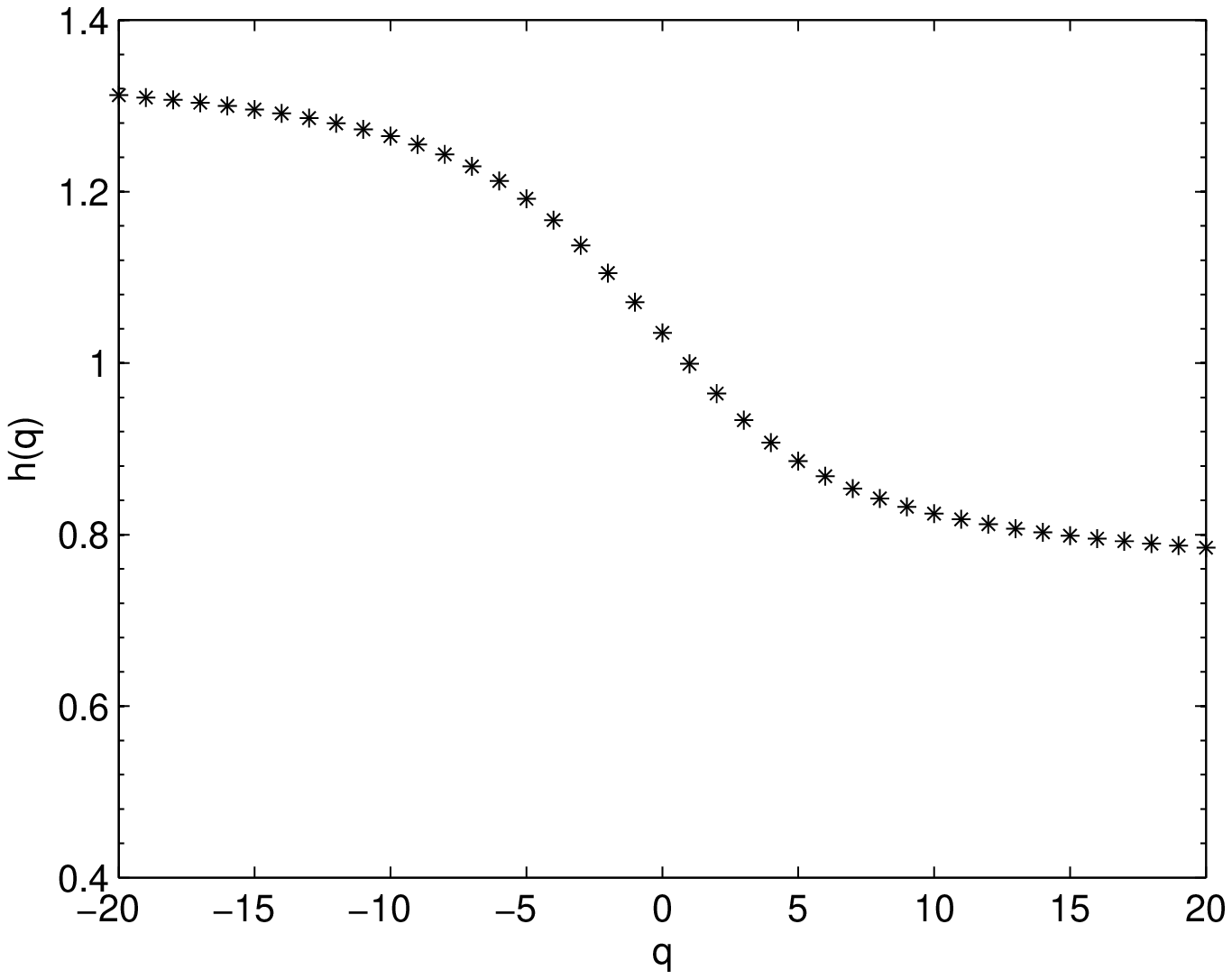}
	  }\\ 
      \end{center}
      \caption{The generalized Hurst exponents as functions of $q$ for Equivalent Ice Extent (EIE) (a) {\em with a seasonal cycle} and (b) {\em without a seasonal cycle}, and the ice albedo (c) {\em with a seasonal cycle} and (d) {\em without a seasonal cycle}.   }%
      \label{fig:Hurst}
\end{figure}

\subsection{Generalized Hurst exponents  \label{sec:Hurst}}

Figure  \ref{fig:Hurst} shows the trends in the generalized Hurst exponents, $h(q)$, associated with the strong seasonal cycle. We calculate these exponents, which describe the slope of the fluctuation curve, for each $q$ as the linear approximation of the entire curve.   It is principally used here to demonstrate the strong influence of the seasonal cycle in suppressing $h(q)$ for all $q$.   There are several important effects immediately evident.  First, the general reduction in the magnitudes of the $h(q)$ associated with the seasonal cycle is synonymous with the suppression of long term persistence, restricting the crossovers to short term dynamics and masking behavior on time scales longer than seasonal, as was seen in the fluctuation functions themselves in   
Figs. \ref{fig:EIEwss} and \ref{fig:albdwss}.  Second, this suppression of long term persistence by definition suppresses multiple crossovers as well as trends on long time scales.  Third, the combination of the reduced magnitude and the general flatness of the curves with the seasonal cycle present indicates a suppression of the ``dynamic range'' of multifractality.  The analysis and interpretation of related multifractal exponents is the subject of a separate publication.  The main point of Fig. \ref{fig:Hurst} is to display the behavior and interpretation of Figs. \ref{fig:EIEwss} and \ref{fig:albdwss} in a more compact manner.  

\subsection{Discussion \label{sec:discuss}}

In his EIE observational analysis \citet[][]{IanGeom} noted that the overall retreat of the ice cover is governed by a noisy signal and particularly so at the September minimum.  Observations using $\sim$ 30 years of monthly ice extent and area data show rapid decorrelations of a few months \cite[][]{BW:2011} using lag correlation method assuming that the underlying process is a first-order autoregressive process.  A similar approach was applied to fit hindcasting simulation output which showed a persistence time scale for September ice area of 1.2 years \cite[][]{Armour:2011}.  In a study of the forecast skill of a linear empirical model \citet[][]{Lindsay:2008} found no skill in predicting detrended data for time scales of three or more months.  Finally, the rapid decay in ice area anomalies is also found in climate models \cite[see][and refs therein]{Armour:2011b} and this is argued to underlie the lack of hysteresis associated with the loss of summer sea ice in both global models and theoretical treatments  such as in \citet[][]{EW09}.

The corpus of these studies lead to the general conclusion of rapidly decaying correlations and hence compromised predictability.  Thus, in this respect they are heuristically consistent with our interpretation here.  However,  there remain important distinctions.
Firstly, even a single moment of the fluctuation function demonstrates the existence of multiple time scales in the data which, as noted in Section \ref{sec:twdfa}\ref{sec:mf}, cannot be treated in a quantitatively consistent manner with a single decay autocorrelation.  Moreover, we showed in Section  \ref{sec:results}\ref{sec:masking} that if the seasonal cycle is not removed one will always observe a single crossover time (longer than the synoptic time scale) of 1 year; a time scale at which all moments converge.  Thus, the upper bound on the persistence time in any study that assumes an autoregressive process will inevitably be $\sim$ 1 year, as is indeed found for ice area \cite[][]{BW:2011, Armour:2011}.   Apart from the single moment evidence of multiple scales in the system, the approach here highlights the dangers of not carefully detrending the seasonality and dealing with stationarity.  Secondly, both data sets as used for this analysis and many of the studies mentioned above do not explicitly distinguish between ice types. Thus, this analysis applies to the aggregate of the ice cover, whereas, subject to a wide range of caveats, many models can diagnose features such as thickness or type.   Present work focuses on extending this analysis in to various ice types.  

\section{Concluding Remarks}

We have examined the long-term correlations and multifractal properties of daily satellite retrievals of Arctic sea ice albedo and equivalent ice extent (EIE), for periods of $\sim$ 23 years and 32 years respectively, with and without the seasonal cycle removed.  A recent development  called Multifractal Temporally Weighted Detrended Fluctuation Analysis (MF-TWDFA), which exploits the intuition that in any time series points closer in time are more likely to be related than distant points, was adapted for use with these data.  Points in the records nearer each other are weighted more than those farther away in order to determine a polynomial used to fit the time series {\em profile}, which is a cumulative sum of the time series.  As a methodology the approach offers several advantages over the more generally applied MF-DFA.  The profile in MF-DFA is fit using discontinuous polynomials, which can introduce errors in the determination of crossover times for new scalings, and can be particularly questionable for long time scales.  
Whereas MF-DFA is typically informative only up to $N/4$ for time series of length $N$, MF-TWDFA can be carried out to $N/2$.  Additionally, the generalized fluctuation functions $F_q (s)$ for all moments $q$ as a function of time scale $s$ are substantially smoother for all $s$ and this is particularly so for large values.  This facilitates clear extraction of crossover times from one scale to another.  

The generalized Hurst exponents and multiple crossover timescales were found to range from the synoptic or weather time scale to decadal, with several between.  Such multiple time scales were exhibited in both data sets and hence the approach provides a framework to examine ice dynamical and thermodynamical responses to climate forcing that goes beyond treatments that assume a process involving a single autocorrelation decay,  such as a first-order autoregressive process.  Indeed, the method shows that single decay autocorrelations cannot be meaningfully fitted to these geophysical observations.  One of our most important findings is that the strength of the seasonal cycle is such that it dominates the power spectrum and ``masks'' long term correlations on time scales beyond seasonal.  When detrending the seasonality from the original record, the EIE data exhibits a white noise behavior from seasonal to bi-seasonal time scales, but the clear fingerprints of the short (weather) and long ($\sim$ 7 and 9 year) time scales remain, demonstrating a {\em reentrant long-term persistence}.  Therefore, it is not possible to distinguish whether a given EIE minimum (maximum) will be followed by a minimum (maximum) that is larger or smaller.  This means that while it is tempting to use an anomalous excursion associated with a low ice year to predict the following year's minimum, or that two year's henceforth, the present data do not justify such a prediction.  

We tested the idea that the rapid decline in the ice coverage during the most recent decade is associated with the longest crossovers found in our analysis of the EIE.  This was done by reanalyzing this record between 1982-2004, the time range corresponding to the albedo data set.  In so doing, we found that the exclusion of the most recent decade led to the two long term crossovers decreasing from 7 to 6.2 and from 9 to 8.4 years.  Moreover, the white noise structure on seasonal to bi-seasonal times is the same in both records.  Hence, the increase in the crossover times associated with including the last decade of data appears to be the most  plausible reflection of the recent declines.  However, we have not attempted to correlate these longer time scales with other climate indices.  

Finally, other methods of analyzing such data find solely a rapid decorrelation, whereas we find multi-year and decadal transitions as well as the origin of the dominance of the seasonal cycle in long term persistence.   Hence, we believe that combining such multifractal studies of model output and other observations will substantially improve the acuity with which one can disentangle the strength of the seasonal cycle in this highly forced system from the longer term trends.

\ack{
WM thanks NASA for a graduate fellowship.
JSW thanks the Wenner-Gren and John Simon Guggenheim Foundations, the Swedish Research Council and Yale University for support.  The authors thank A.J. Wells and the referees for their comments and suggestions.  
}


\end{document}